\documentclass[journal=jpccck,manuscript=article]{achemso}

\usepackage[T1]{fontenc} 

\usepackage{hyperref}

\author{Nikita S. Pavlov}
\affiliation{Institute for Electrophysics, Russian Academy of Sciences, Ekaterinburg 620016, Russia}
\email{pavlov@iep.uran.ru}

\author{Timur K. Kim}
\affiliation{Diamond Light Source, Harwell Campus, Didcot, OX11 0DE, United Kingdomk}

\author{Alexander Yaresko}
\affiliation{Max Planck Institute for Solid State Research, Heisenbergstra{\ss}e 1, 70569 Stuttgart, Germany}

\author{Ki-Young Choi}
\affiliation{Agency for Defense Development Convergence Technology Collaboration Directorate Yuseong P.O. Box 35, Daejon,34186, Korea}

\author{Igor A. Nekrasov}
\affiliation{Institute for Electrophysics, Russian Academy of Sciences, Ekaterinburg 620016, Russia}

\author{Daniil V. Evtushinsky}
\email{daniil.yevtushynsky@epfl.ch}
\affiliation{Laboratory for Quantum Magnetism, Institute of Physics, Ecole Polytechnique F\'{e}derale de Lausanne (EPFL), CH-1015 Lausanne, Switzerland}

\title{Weakness of Correlation Effect Manifestation in BaNi$_2$As$_2$: ARPES and LDA+DMFT Study}

\abbreviations{ARPES,DMFT,DFT}

\begin{document}

\begin{tocentry}
		\includegraphics[width=1.0\linewidth]{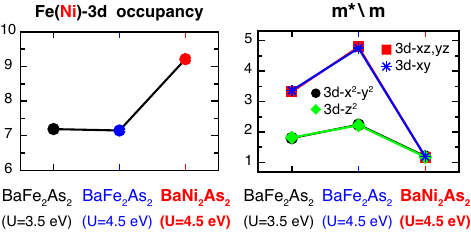}
\end{tocentry}

\begin{abstract}
The electronic spectral function of BaNi$_2$As$_2$ is investigated using both the angle-resolved photoemission spectroscopy (ARPES) and a combined computational scheme of local density approximation together with dynamical mean-field theory (LDA+DMFT).
In contrast to well studied isostructural iron arsenide high temperature superconductors, the BaNi$_2$As$_2$ demonstrate weak correlation effects although Ni-3d elections have even lager on-site interaction than Fe-3d ones.
LDA+DMFT effective mass enhancement for bands crossing the Fermi level is found to be only about~$1.2$ which agrees well with ARPES data.
This reduction of the correlation manifestation with respect to iron pnictides comes from the increase of 3d-orbital filling, when going from Fe to Ni.
The electron correlations cause remarkable reconstruction of the bare BaNi$_2$As$_2$ LDA band structure below $-0.8$~eV due to self-energy effect.
A simplified toy model to understand weakness of correlation effects in BaNi$_2$As$_2$ and to describe the LDA+DMFT self-energy shape is discussed.
For more realistic comparison of LDA+DMFT spectral function maps with ARPES data we take into account several experimental features: the photoemission cross-section, the experimental energy and angular resolutions and the photo-hole lifetime effects.
Thus presented here LDA+DMFT calculations with experimental features included provide nearly qualitative agreement with ARPES data and assure the observation of a dramatic apparent decrease of the correlation strength compared to the Fe compounds.
\end{abstract}

\section{Introduction}
At present, for more than ten years, new high-temperature iron-based superconductors are being intensively studied both experimentally and
theoretically
(cf. reviews~\cite{Sad_08, Hoso_09, John, MazKor, Stew, Kordyuk_12}). Rather soon after the discovery of high temperature superconductivity in
iron pnictides it was realized that electronic correlations are quite significant there. Electron-electron interactions beyond the mean-field
approach are one of the most complicated and promising phenomena in modern condensed matter
physics~\cite{LDADMFT_Held2001,LDADMFT_Kotliar2006}, and often the description of such interacting systems via the single particle self-energy
turns out to be very useful.
The combined computational scheme of the local density approximation and dynamical mean-field theory
(LDA+DMFT)~\cite{LDADMFT_Held2001,LDADMFT_Kotliar2006} is a versatile tool allowing to address the spectrum of the systems with various levels
of the electronic correlation strength from the theoretical point of view. The experimental counterpart, providing direct access to the same
electron spectral function, is the angle-resolved photoemission spectroscopy (ARPES)~\cite{Damascelli_ARPES_2003}. Very important for the
electronic properties of a material, and widely discussed manifestation of the many-body effects is the renormalisation of the band dispersion
at the Fermi level. Thus, it is already well known that in the iron-based superconductors, for the most bands crossing the Fermi level, the
effective mass renormalization is about $\sim 3$ for '11' crystalline type, FeSe it is estimated as 3 - 5, for '111'\,---\,2 - 3.5, for
'1111'\,---\,2 - 3, for '122'\,---\,2 - 3, for $A_x$Fe$_2$Se$_2$\,---\,2 - 3~\cite{Haule2008,Anisimov2009,Yin2011}. Moreover the role of Hund's
coupling alone was widely discussed for iron-based superconductors~\cite{Haule2009,DeMedici2011}. Because of nearly half-filled Fe-3d shell the
critical interaction strength $U_c$ (where Mott transition occurs) and consequently quasiparticle renormalization factor $Z$ are decreasing with $J$ increasing.

As it turned out, one can synthesize many materials isostructural to the iron superconductors. For example, once arsenic is completely
substituted by selenium~\cite{FeSe}, the superconductivity with high critical temperatures can still be achieved, and the level of the
correlation strength in the electronic system seems to persist~\cite{PvsC, JTLRev}. Taking into account that so far everything points towards
the fact that the most important element in the structure of the iron superconductors is actually iron, it becomes very interesting to look at
what happens to the electronic properties once we substitute it. We consider the case of the BaNi$_2$As$_2$\,---\,one can see it as a result of
swapping iron by nickel in the prototypical compound BaFe$_2$As$_2$. From the very beginning, it was pointed out that the band renormalization
in BaNi$_2$As$_2$ at the Fermi level is much weaker than in iron pnictides~\cite{BaNi2As2_ARPES_2011,BaNi2As2_LDA_Chen_PRB_2009}. However, to
our knowledge, electron correlation effects in BaNi$_2$As$_2$ were not analyzed in details. Thus, there is an open question why manifestation of the correlations in BaNi$_2$As$_2$ is rather weak~\cite{BaNi2As2_ARPES_2011} in contrast to the analogous superconducting iron pnictides?

In this paper we present a joint experimental-theoretical study of BaNi$_2$As$_2$ using ARPES measurements and LDA+DMFT of the electronic spectrum on the scales of the entire Ni-3d and As-4p bands.
We explicitly demonstrate that the band renormalization is rather weak near the Fermi level in BaNi$_2$As$_2$ because of almost occupied Ni-3d states.
However correlation effects in BaNi$_2$As$_2$ remarkably reconstruct bare LDA bands at the higher binding energies due to self-energy effects.  Finally in Appendix a simplified toy model to understand weakness of correlation effects in BaNi$_2$As$_2$ and to describe the LDA+DMFT self-energy shape is proposed.

To provide a better comparison of LDA+DMFT spectral functions with ARPES data the photoemission cross-section, the experimental energy and
angular resolutions and the
photo-hole lifetime effects are included into consideration (details of that are given in Appendix).
All these give good, almost quantitative agreement with experimental photoemission data.

\subsection{Previous works on BaNi$_2$As$_2$}

Despite the popularity and importance of the BaFe$_2$As$_2$ compound in the studies of the unconventional superconductivity and relevant
effects, the BaNi$_2$As$_2$ system has been addressed less widely.
In the BaNi$_2$As$_2$ there is no spin density wave (SDW) or any other type of magnetic ordering in contrast to iron arsenides~\cite{BaNi2As2_ARPES_2011}.
At the same time, there is a structural transition, and, as suggested in Ref.~\cite{Lee2019}, the charge density wave (CDW) might occur at low temperatures.
The low-temperature evolution of thermal conductivity and specific heat in BaNi$_2$As$_2$ is quite similar to the conventional BCS superconductors~\cite{BaNi2As2_k_C_PRL_2009}.
The behavior corresponds to a weak coupling regime with $T_c = 0.7$~K~\cite{BaNi2As2_struct_all_2008}, which is in good agreement with the simple BCS-like $T_c$ estimation based on DFT/LDA calculated density of states (DOS) on the Fermi level~\cite{deltaZ} done in Refs.~\cite{PvsC,SrPtAs}.
A number of LDA calculations were done for BaNi$_2$As$_2$~\cite{BaNi2As2_ARPES_2011,
BaNi2As2_LDA_Chen_PRB_2009,BaNi2As2_LDA_ShI_2009,BaNi2As2_LDA_Chen_2010}.

There are a couple of ARPES studies of the BaNi$_2$As$_2$~\cite{BaNi2As2_ARPES_2011,Noda2017}.
In the early work~\cite{BaNi2As2_ARPES_2011} it was reported weakness of correlation effects on the Fermi level for both triclinic and tetragonal phases via comparison of ARPES and DFT bands.
In the work of Noda $et~al.$~\cite{Noda2017} ARPES data for both phases is juxtaposed with band calculations from~\cite{BaNi2As2_LDA_Chen_PRB_2009} by shifting and scaling the individual branches of the DFT dispersions independently in the spirit of Kordyuk's work~\cite{Kordyuk_12} to discuss renormalization effects near the Fermi level as well.
Some speculations (according to the authors of~\cite{Noda2017}) on the Peierls transition scenario to understand the reasons of the band structure change because of the tetragonal to triclinic phase transition were introduced.
However, the articles~\cite{BaNi2As2_ARPES_2011,Noda2017} do not discuss at all the initial reason of the weak manifestation of correlation effects in both phases of BaNi$_2$As$_2$.

\section{Computational Details}

The crystal structure of BaNi$_2$As$_2$ above $T=131$ K has the tetragonal symmetry with I4/mmm space group.
Lattice constants are $a=b=4.112$~\AA\ and $c=11.542$~\AA.
The atomic positions in the elementary unit cell are following:
Ba --- 2a($0,0,0$), Ni --- 4d($0,0.5,0.25$), As --- 4e($0,0,0.3476$)~\cite{BaNi2As2_struct_all_2008}.
As it shown earlier~\cite{BaNi2As2_ARPES_2011} the band structure of triclinic phase is very similar to that of tetragonal phase (with the difference of band positions not more than $0.1-0.2$ eV, or even less, in the directions where our ARPES is measured).
Most pronounced difference between tetragonal and triclinic bands appears in the middle of (0,0,0)-(0.5,0,0) direction is about 0.4 eV because of the allowance of Ni-3d$_{xy}$ -- Ni-3d$_{x^2-y^2}$ hybridizion channel with symmetry lowering.
While the total and partial DOSes in the triclinic and tetragonal phases are nearly identical.
So we chose the tetragonal structure.
The LDA calculations were performed within the Elk full-potential linearized augmented plane-wave (FP-LAPW) code~\cite{ELK}.
We employed the CT-QMC impurity solver~\cite{ctqmc_PRL_2006,ctqmc_PRB_2007,ctqmc_RMP_2011,triqs_1,triqs_2,triqs_3_leg} for the DMFT part of
LDA+DMFT calculations.
To define the DMFT lattice problem for BaNi$_2$As$_2$ compound we used the LDA Hamiltonian projected to Ni-3d, As-4p and Ba-5d Wannier functions. The Wannier projection works very well here, and the Wannier-projected dispersions perfectly coincide with the LDA ones below $+2$~eV.
Wannier functions are obtained within exciting-plus code.

The DMFT(CT-QMC) computations were done at the inverse temperature $\beta=40$ ($\sim290$ K)
with about $10^8$ Monte-Carlo sweeps.
The interaction parameters of the Hubbard model were taken as $U=4.5$ eV and $J=0.85$ eV.
The $U$ value is slightly larger~\cite{Solovyev1994} than typical values for iron arsenides~\cite{BaFe2As2_Skornyakov2012,BaFe2As2_Werner2012,NaFeAs}.
The $U$ matrix was taken in 2-index form~\cite{FLLpaper2} (reduced from full 4-index matrix~\cite{Lichtenstein_Katsnelson98}) which corresponds only to density-density terms in the Hamiltonian.
Such choice of the Hamiltonian and U matrix parametrization is typical for multi-orbital cases e.g. iron-based superconductors~\cite{BaFe2As2_Skornyakov2012,LiFeAs_Ferber2012}.
We employed the self-consistent fully-localized limit definition of the double-counting correction ($E_{dc}$)~\cite{FLLpaper1,FLLpaper2}.
Thus computed values of the Ni-3d occupancies and the corresponding double-counting energies are $n_d=9.21$, $E_{dc}=35.69$ eV.
The LDA+DMFT spectral function maps were obtained after an analytic continuation of the local self-energy $\Sigma(\omega)$ from the Matsubara frequencies to the real ones. To this end we have applied the Pade approximant algorithm~\cite{pade} and checked the results with the maximum entropy method~\cite{ME} for the Green's function G($\tau$).

The crystal structure of BaFe$_2$As$_2$ has I4/mmm space group with $a=b=3.909$~\AA\ and $c=13.212$~\AA,
As --- 4e($0,0,0.3538$)~\cite{Rotter2008}. And all other computational details are the same except for the Wannier functions chosen -- Fe-3d, As-4p.
The LDA+DMFT computed values of the Fe-3d occupancies and the corresponding double-counting
energies are $n_d=7.18$, $E_{dc}=20.74$ eV for $U=3.5$ eV, $J=0.85$ eV and
$n_d=7.14$, $E_{dc}=27.25$ eV for $U=4.5$ eV, $J=0.85$ eV.

\section{Experimental Details}
The ARPES experiments were performed at the I05 beamline of the Diamond Light Source~\cite{Hoesh2017}.
The single crystalline samples were cleaved in ultra high vacuum (better than 2$^{-10}$ mbar).
The photoemission spectra with total energy resolution of 20~meV and angular resolution of 0.5$^{\circ}$ were measured at 7~K, using photons with the energy 120~eV of various polarizations.
In order to increase the contrast of second derivative of ARPES here we have used a procedure similar to the one described in~\cite{ARPES_second_deriv}.

Experimental density of state was obtained by integrating the photoemission intensity distribution recorded at the photon energy of 120\,eV along the cut passing through the Brillouin zone diagonal.

\section{Results and Discussion}

\begin{figure}[h]
	\includegraphics[width=0.49\linewidth]{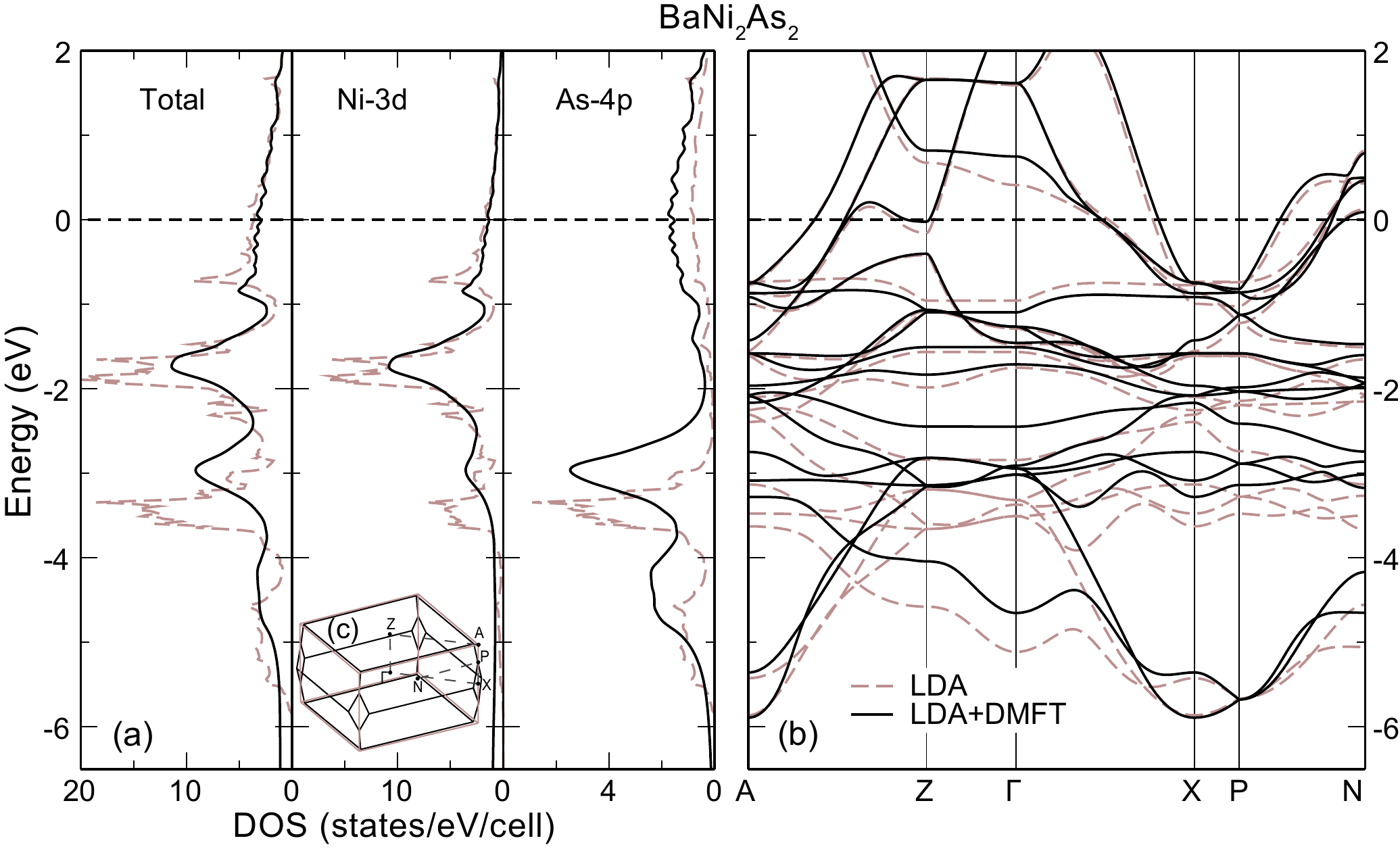}
	\includegraphics[width=0.47\linewidth]{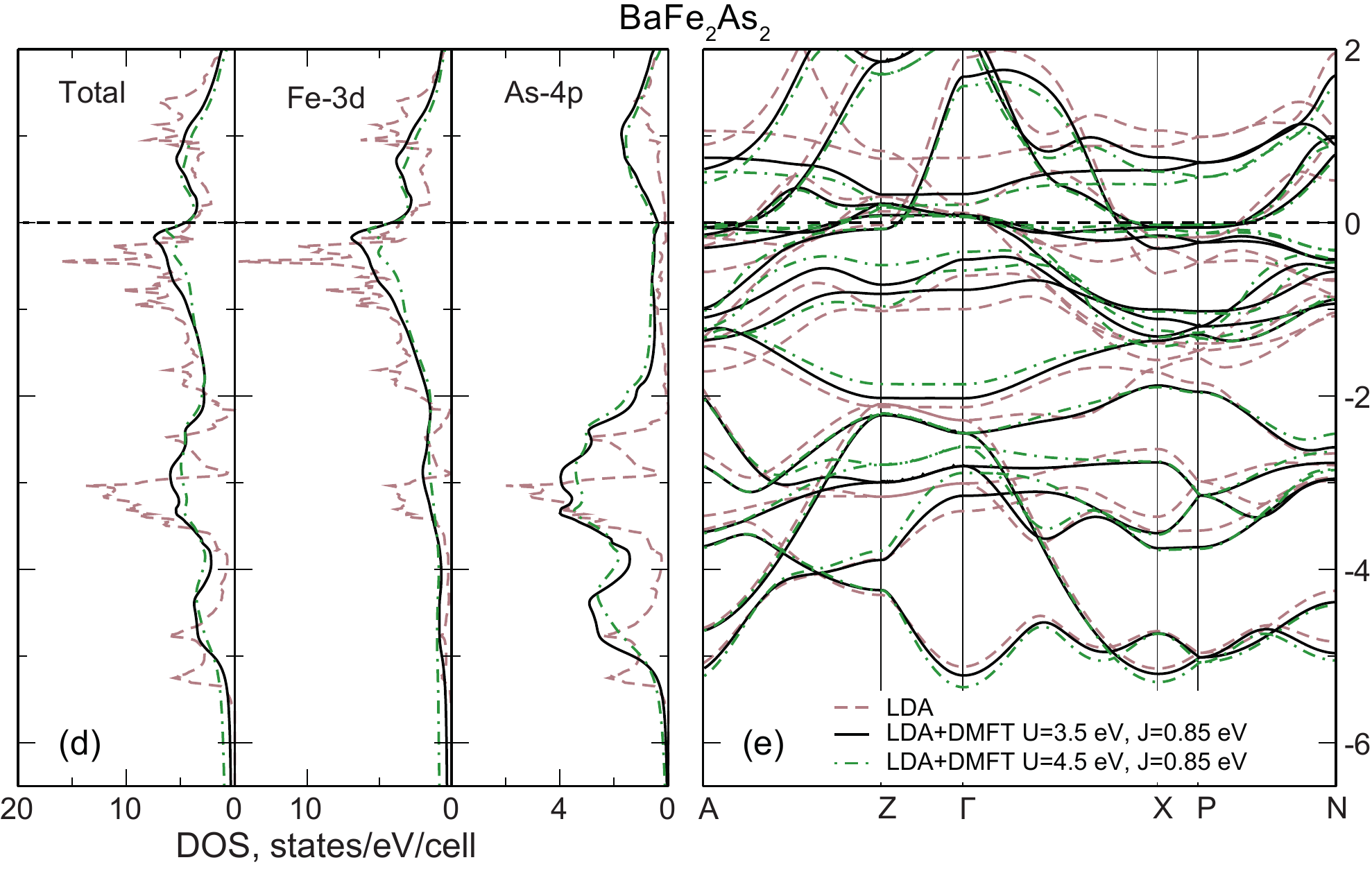}
	\caption{LDA (dashed lines) and LDA+DMFT (solid lines) calculated densities of states (a) and band dispersions (b)
		of paramagnetic BaNi$_2$As$_2$ and BaFe$_2$As$_2$ --- panels (d) and (e) correspondingly.
		On the panel (e) LDA+DMFT bands obtained for typical iron arsenides interaction parameters (solid lines)  are compared with those for intaration strength of BaNi$_2$As$_2$ (dot-dashed lines).
		Panel (c) --- body-centred tetragonal first Brillouin zone (black lines);  primitive tetragonal Brillouin zone (brown lines); k-path for band dispersions is shown with dashed lines.
		A-point corresponds to ($\pi/a, \pi/a, \pi/c$). The Fermi level is at zero energy.}
	\label{LDA_dos_bands}
\end{figure}
In Figure~\ref{LDA_dos_bands} LDA and LDA+DMFT calculated total, Ni-3d and As-4p densities of states (panel (a)) and band dispersions (panel (b)) for the BaNi$_2$As$_2$ are shown.
First of all bare LDA band structure of BaNi$_2$As$_2$ is quite different to a typical one of iron arsenides {\it e.g.} BaFe$_2$As$_2$ (see panels (d) and (e) of Figure~\ref{LDA_dos_bands}).
There are three bands crossing the Fermi level in the center of $\Gamma$-X direction~\cite{BaNi2As2_LDA_ShI_2009} instead of five in BaFe$_2$As$_2$~\cite{Nekrasov2008} (three near $\Gamma$-point and two near X-point) and four instead of two in the P-N direction respectively.
Thus Fermi surface of BaNi$_2$As$_2$ does not have hole cylinders near $\Gamma$-point~\cite{BaNi2As2_LDA_ShI_2009} in contrast to BaFe$_2$As$_2$~\cite{Nekrasov2008}.
Near X-point BaNi$_2$As$_2$ has three cylinders~\cite{BaNi2As2_LDA_ShI_2009} while BaFe$_2$As$_2$ has just two of more or less the same shape.
However one should note that band composition of the Fermi surface sheets is totally different in BaNi$_2$As$_2$~\cite{BaNi2As2_LDA_ShI_2009}
and BaFe$_2$As$_2$~\cite{Nekrasov2008}.
Despite the fact that both systems are isostructural to each other, formally the Ni-compound has four more electrons per formula unit. However, the BaFe$_2$A$_2$ bands cannot be obtained by a rigid band shift of the corresponding BaNi$_2$As$_2$ bands (compare panels (b) and (e) in Figure~\ref{LDA_dos_bands}).
Moreover, if we look at the densities of Fe(Ni)-3d states, we will see that they to some extent resemble each other up to a shift of 1.5~eV.
The As-4p states, which were at $1$~eV in BaFe$_2$As$_2$, got to the Fermi level in BaNi$_2$As$_2$.
However, the hybridization of the Fe(Ni)-3d -- As-4p states did not change significantly (Ni-3d bandwidth is only about 5\% larger than Fe-3d one).
But, there is spectral weight shift of As-4p states upwards in energy induced by correlations on Ni-3d states for BaNi$_2$As$_2$. For BaFe$_2$As$_2$ this effect is absent.

It is well established that direct Coulomb electron-electron interaction and Hund's coupling are important for iron based superconductors~\cite{Haule2008,Anisimov2009,Yin2011,DeMedici2014}.
The Ni-3d elections in BaNi$_2$As$_2$ should have even lager on-site interaction in comparison to Fe-3d ones~\cite{Solovyev1994}.
To this end we use LDA+DMFT method to account for the local Coulomb interactions in BaNi$_2$As$_2$.
In contrast to well studied isostructural iron arsenide high temperature superconductors ($e.g.$~BaFe$_2$As$_2$), the BaNi$_2$As$_2$
demonstrate weak correlation effects manifestation with effective mass enhancement for bands crossing the Fermi level only about~$1.2$ without any orbital dependence (see Table~\ref{Tab1}).
This reduction of the correlation manifestation with respect to iron pnictides comes from the increase of 3d-orbital filling, when going from Fe to Ni (see following discussion).

\begin{table}
	\caption{The effective mass enhancement ($Z^{-1} = 1-\rm{Im}(\Sigma(i \omega_0))/\omega_0$) for different systems and interaction parameters.}
	\begin{tabular}{|l|cccc|}
	\hline					
& 3d$_{x^2-y^2}$ & 3d$_{yz,xz}$ & 3d$_{z^2}$ & 3d$_{xy}$ \\
	\hline
BaNi$_2$As$_2$ (U=4.5, J=0.85 eV)
&       1.20     &      1.18    &     1.18   &   1.19 \\
BaFe$_2$As$_2$ (U=3.5, J=0.85 eV)
&       1.80     &      3.32    &     1.82   &   3.36 \\
BaFe$_2$As$_2$ (U=4.5, J=0.85 eV)
&       2.24     &      4.79    &     2.22   &   4.75 \\
	\hline
	\end{tabular}
	\label{Tab1}
\end{table}

Thus, the LDA+DMFT obtained bands (solid lines on Figure~\ref{LDA_dos_bands}(b)) are almost identical near the Fermi level to the LDA ones (dashed lines on Figure~\ref{LDA_dos_bands}(b)). Similar picture is obtained for densities of states (Figure~\ref{LDA_dos_bands}(a)).
This fact was experimentally demonstrated by ARPES in the work~\cite{BaNi2As2_ARPES_2011} where near the Fermi level renormalization factor was found to be about 1.6.
However bands below $-0.8$ eV are noticeably modified by correlations within LDA+DMFT (solid lines on Figure~\ref{LDA_dos_bands}(b)) in comparison to LDA bands (dashed lines on Figure~\ref{LDA_dos_bands}(b)).
The positions of LDA+DMFT bands located in interval from $-0.8$ to $-5$ eV have non uniform shift upto $0.5$ eV by the correlations with respect to LDA ones.

For clarity of comparison the LDA+DMFT calculations of BaFe$_2$As$_2$ system are given (panels (d) and (e) of Figure~\ref{LDA_dos_bands}).
For conventional values of interaction in BaFe$_2$As$_2$ (U=$3.5$ eV, J=$0.85$ eV) the mass renormalization is from 1.8 to 3.4 for different orbitals, which agrees well with many other previous results~\cite{Yin2011,BaFe2As2_Skornyakov2012,BaFe2As2_Werner2012}.
For the same parameters as for BaNi$_2$As$_2$ (U=$4.5$ eV, J=$0.85$ eV) the mass renormalization in BaFe$_2$As$_2$ is quite large  --- from 2.2 to 4.8 for different orbitals (see Table~\ref{Tab1}) with more pronounced orbital dependence than in the  U=$3.5$ eV, J=$0.85$ eV case.

\begin{figure}[h]
	\includegraphics[width=0.4\linewidth]{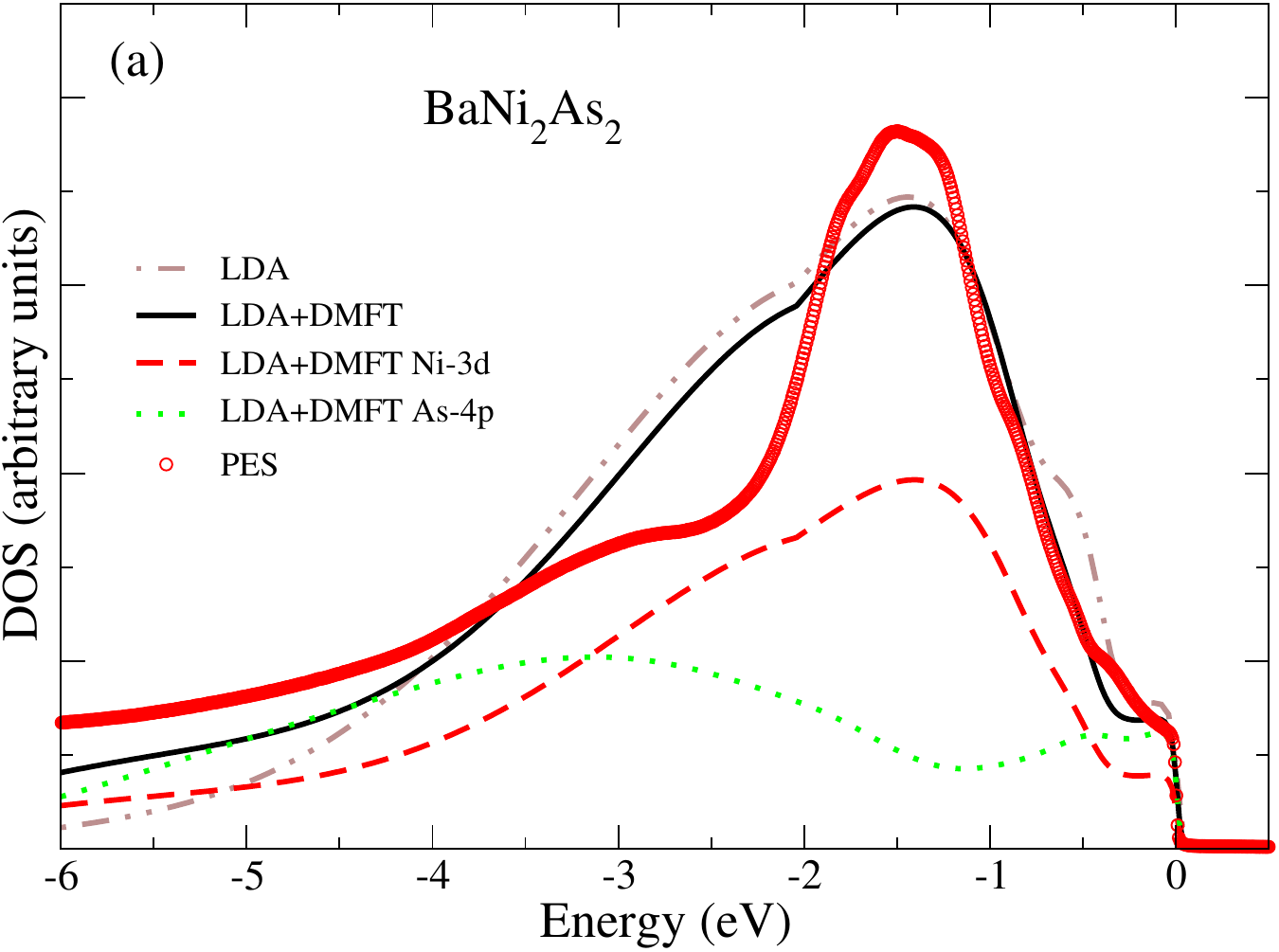}
	\includegraphics[width=0.4\linewidth]{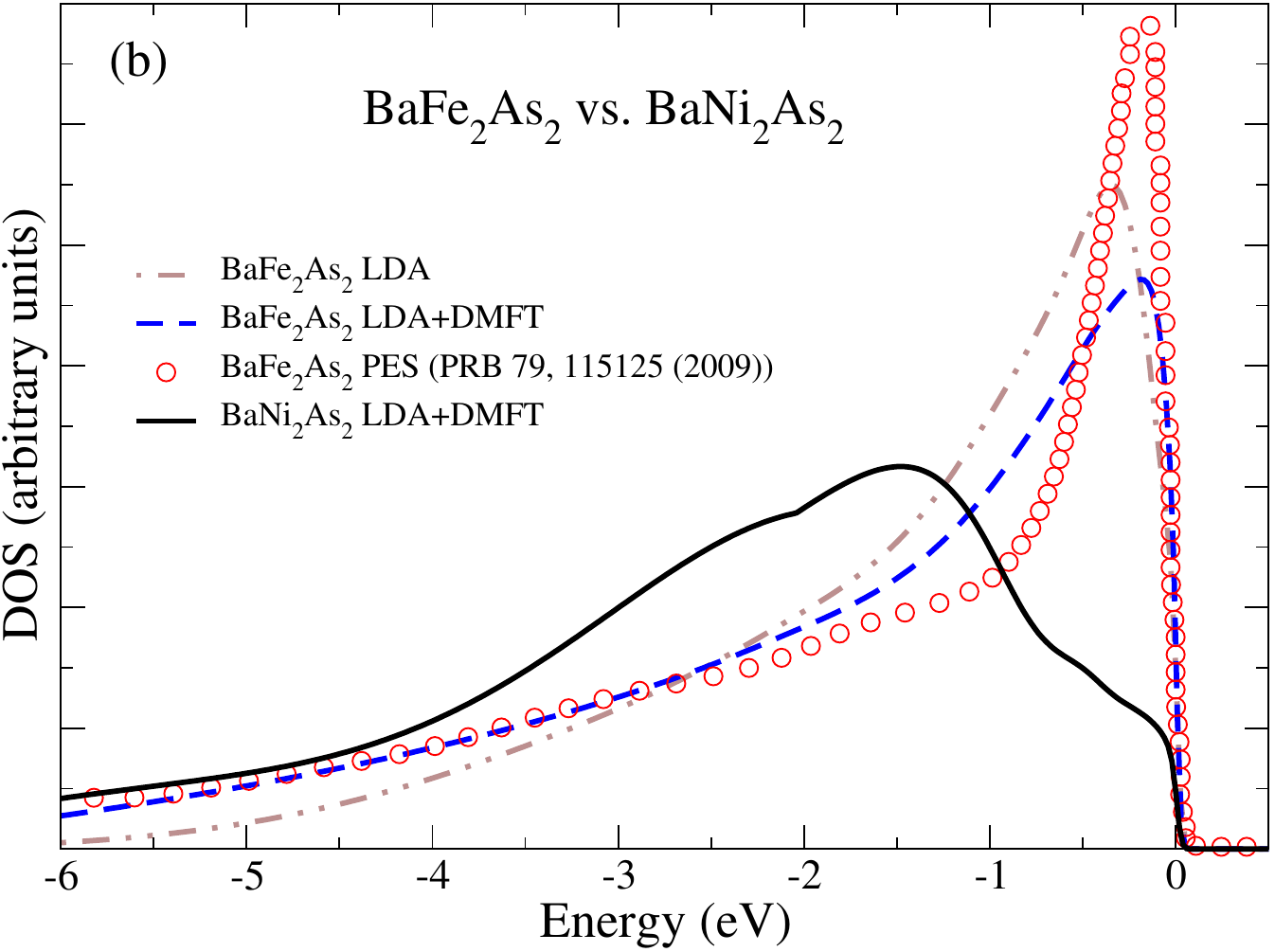}
	\caption{Panel (a): Photoemission spectrum (PES) of BaNi$_2$As$_2$ obtained by integration of a unitary ARPES image is shown as red circles.
		Theoretical PES: LDA (dash-dotted brown line), LDA+DMFT (black solid line), LDA+DMFT Ni-3d states only (red dashed line), LDA+DMFT As-4p states only (green dotted line).
		Panel (b): PES of BaFe$_2$As$_2$ from~\cite{DeJong2009} (red circles), LDA (dash-dotted brown line), LDA+DMFT (blue dashed line), LDA+DMFT BaNi$_2$As$_2$ (black solid line).
		The Fermi level is at zero energy.}
	\label{PES}
\end{figure}

First of all we compare experimental photoemission spectrum to LDA or LDA+DMFT calculated spectra (Figure~\ref{PES} panel (a)).
To compare theory and experiment one needs to take into account the photoemission cross-section, the energy resolution and the photo-hole lifetime effects (see details on that in Appendix).

The position of the main peak (about $-1.5$~eV) agrees well between experimental and theoretical PES.
The electronic states, yielding the most intensity here, belong to the bands of the predominant Ni-3d$_{z^2}$ and Ni-3d$_{yz,xz}$ character.
Peak in PES about $-3.0$~eV is formed by Ni-3d and As-4p states.
There is small difference between non-correlated (LDA) and correlated (LDA+DMFT) system in the PES. 
The BaNi$_2$As$_2$ system has wider and flater PES than BaFe$_2$As$_2$ system (Figure~\ref{PES} panel (b)).
The BaFe$_2$As$_2$ PES has main contribution near Fermi level (sharp peak).
Therefore, the BaNi$_2$As$_2$ and BaFe$_2$As$_2$ systems are significantly different.
And we will see that in ARPES too.
Possible reasons why the intensity of the $-3$~eV for BaNi$_2$As$_2$ and $-1.5$~eV BaFe$_2$As$_2$ peaks are overestimated for the theoretical PES might be explained by energy and polarization dependences of matrix elements and/or orbital character dependence of matrix elements.


\begin{figure}[!h]
	\includegraphics[width=0.9\linewidth]{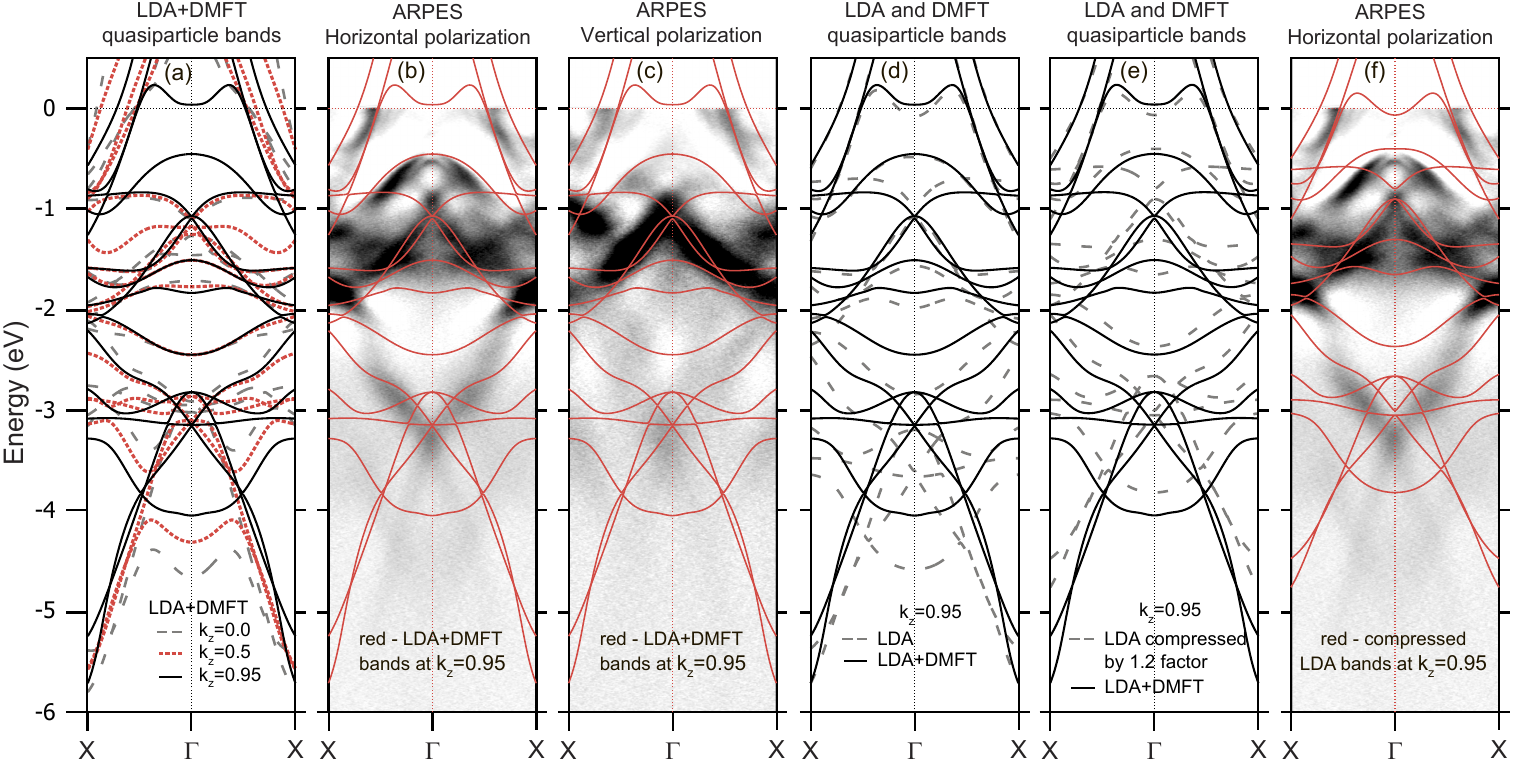}
	\caption{Comparison of LDA+DMFT quasiparticle bands with different $k_z$ vales in units of $\pi/c$ (panel (a)).  Comparison of ARPES data
		second derivatives BaNi$_2$As$_2$ for horizontal (b) and vertical (c) polarizations with LDA+DMFT quasiparticle bands (red lines) and with LDA
		bands (red lines) (f) with $k_z=0.95\ \pi/c$. LDA+DMFT and LDA (d), LDA+DMFT and compressed LDA (e) quasiparticle bands with $k_z=0.95\ \pi/c$.
		The Fermi level is at zero energy. ARPES data recorded at the photon energy of $120$~eV.}
	\label{ARPES_LDADMFT_bands}
\end{figure}
It is well known that by varying the beam energy in ARPES experiments one can access different values of $k_z$ component of the momenta~\cite{Damascelli_ARPES_2003}.
However to obtain a precise value of $k_z$ one should know an exact geometry of ARPES experiment~\cite{Damascelli_ARPES_2003}, the work function and the inner potential for this particular material~\cite{Brouet_Febased_2012}.
So it is much easier to find the $k_z$ value for given ARPES experiment setup from direct comparison of the second derivatives of ARPES data and the theoretical LDA+DMFT calculated bands at different $k_z$ values.
Evolution of the LDA+DMFT quasiparticle bands with $k_z$ changing is presented in Figure~\ref{ARPES_LDADMFT_bands}(a).

In the panels (b) and (c) of Figure~\ref{ARPES_LDADMFT_bands} we present the LDA+DMFT results (red lines) for BaNi$_2$As$_2$ compared with the ARPES data second derivatives with vertical and horizontal polarizations in the X-$\Gamma$-X direction.
In the theoretical data the quasiparticle band near $\Gamma$-point around $-0.5$ eV (clearly seen in ARPES) is present only at $k_z\approx\pi/c$ (here $c$ is the lattice constant along the z-axis mentioned above).
At $k_z=0.9\ \pi/c$ this band changes shape and looses similarity to the ARPES quasiparticle band.
At $k_z=\pi/c$ the band lays slightly above the experimental one.
Thus we firmly believe that $k_z=0.95\ \pi/c$ provides the best possible agreement of ARPES and LDA+DMFT bands.
For all following figures of this paper we will use $k_z=0.95\ \pi/c$ and thus the notation of the $\Gamma$-point is (0,0, $0.95\ \pi/c$) and of the X-point --- ($\pi/a, \pi/a, 0.95\ \pi/c$).


To illustrate the difference between LDA (dashed lines) and LDA+DMFT (solid lines), the quasiparticle bands in X-$\Gamma$-X direction are ploted on Figure~\ref{ARPES_LDADMFT_bands} (d).
Similar to Figure~\ref{LDA_dos_bands}(b), here we see that at the Fermi level LDA and LDA+DMFT bands do not differ much --- thus correlation effect manifestation  at the Fermi level can be recognized as weak.
Main changes due to correlations happen below $-0.8$ eV, where the quasiparticle bands positions are different upto about $0.5$ eV.

In case of the weak manifestation of correlation effects, one might expect that LDA+DMFT bands are just rescaled LDA bands.
However, in case of multiorbital system there is not only rescaling~\cite{Kordyuk_12}.
The LDA bands compressed $1.2$ times are shown on Figure~\ref{ARPES_LDADMFT_bands} (e,f).
At the Fermi level and down to $-0.8$ eV LDA, compressed LDA and LDA+DMFT quasiparticle bands almost coincide.
Nevertheless the statement that LDA only can describe ARPES data is not correct, because the Ni-3d states must be treated with correlations included since $U/W = 4.5$ eV$/4.75$ eV $\sim 1$.
Such a weak manifestation of correlation effects in BaNi$_2$As$_2$, despite large interaction parameters, is caused by an almost filled Ni-3d shell (see discussion below).


In oder to compare ARPES and LDA+DMFT calculated spectral function maps directly, we consider the following experimental features: the photoemission cross-section, the experimental energy and angular resolutions and the photo-hole lifetime effects (see Figure~\ref{LDA_DMFT_ARPES}), as suggested in~\cite{DMFT_broading}. Particular details on that are given in the Appendix.

\begin{figure*}[!ht]
	\includegraphics[width=0.90\linewidth]{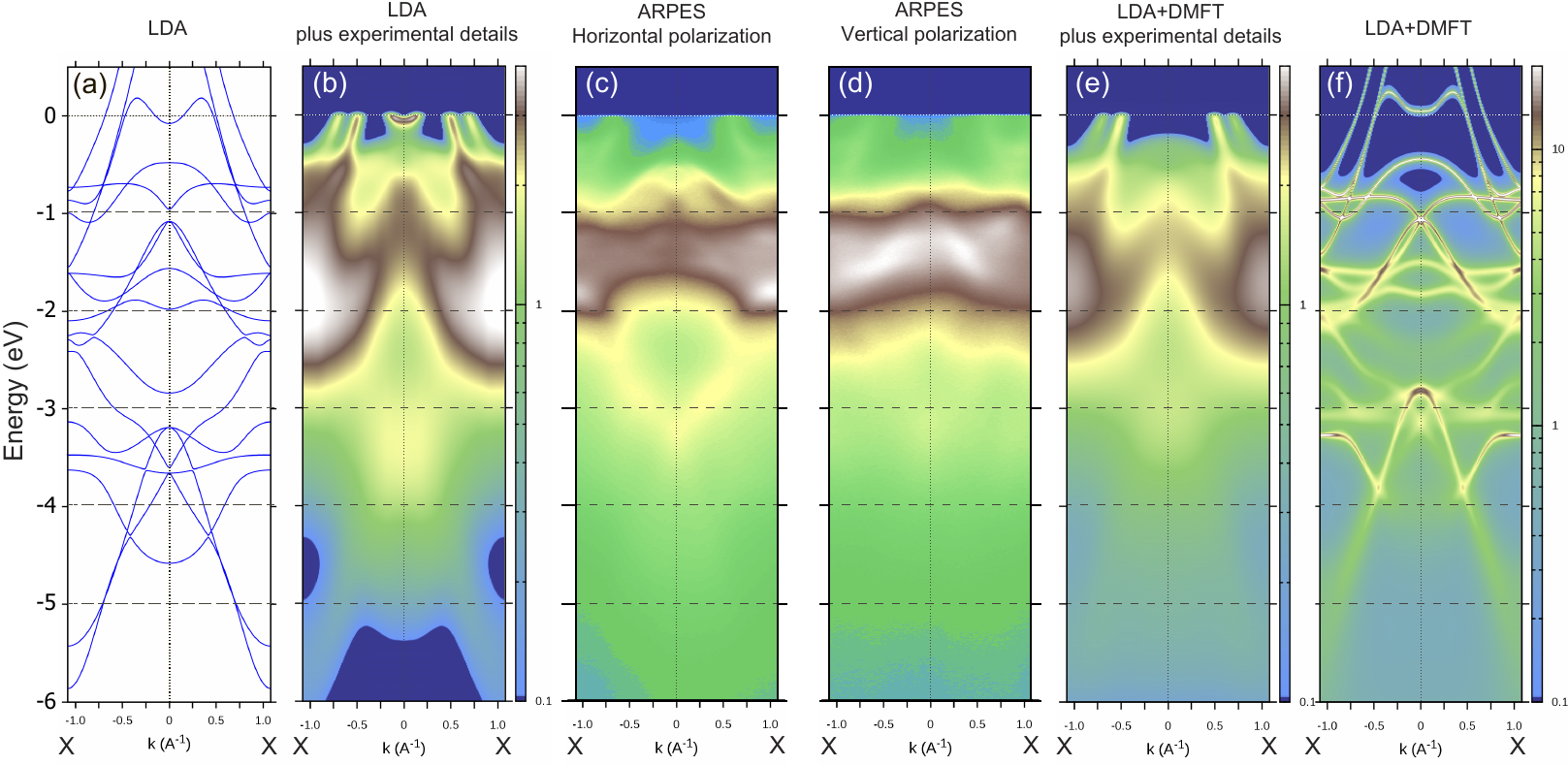}
	\caption{LDA bands (a) and LDA+DMFT (f) spectral function map for BaNi$_2$As$_2$ in the X-$\Gamma$-X high symmetry direction.
		Panels (b) and (e) show LDA and LDA+DMFT spectral function maps with experimental details considered.
		Corresponding LDA and LDA+DMFT spectral functions have the same intensity scale.
		Experimental ARPES data for horizontal (c) and vertical (d) polarizations is given in the middle. The Fermi level is at zero energy.}
	\label{LDA_DMFT_ARPES}
\end{figure*}

In Figure~\ref{LDA_DMFT_ARPES} the LDA (panel (b)) and LDA+DMFT (panel (e)) experimental-like broadened spectral functions are presented together
with the ARPES data (panels (c,d)).
At first glance, LDA spectral function map on Figure~\ref{LDA_DMFT_ARPES}(b) is very similar
to the LDA+DMFT one on Figure~\ref{LDA_DMFT_ARPES}(e).
Mainly, the presence of correlations shifts the bright region near $\Gamma$-point from $-3.5$ eV (LDA) to $-3$ eV (LDA+DMFT).
As a result, despite the weakness of correlation effect manifestation in BaNi$_2$As$_2$, the LDA+DMFT calculations provide better agreement
with experimental ARPES data and almost quantitatively (intensity and position) describes it.
Thus modifications of quasiparticle bands around $-3$ eV can be described as a LDA+DMFT self-energy effect in the tetragonal phase without
involving Peierls scenario~\cite{Noda2017} which is not appropriate for the tetragonal case at all.
In Appendix we propose a simplified toy model to understand weakness of correlation effects and to describe the LDA+DMFT self-energy shape.


Now it is time to discuss the origin of the mentioned weakness of the correlation effects here.
Corresponding 5-band model calculations were done e.g. in
the work of L.~Medici and co-authors in Ref.~\cite{Medici_book}. There it was shown that iron arsenide superconductors with about 6 electrons
per Fe atom and Coulomb interaction about 4.5 eV should have quasiparticle mass renormalization $m^*/m \sim 3$.
For the case of BaNi$_2$As$_2$ our LDA+DMFT results give  Ni-3d state occupancy $\sim 9.2$ electrons which is far away from the situation in
iron arsenides.
Following the model calculations from Ref.~\cite{Medici_book}, for such high filling  and approximately the same Coulomb interaction, the
corresponding mass renormalization should be slightly above one. The same mass renormalization is obtained in our material specific LDA+DMFT
calculations for BaNi$_2$As$_2$.

Another possible reason of that small mass enhancement might be quite large Ni-3d state bare bandwidth ($W$) --- $4.75$ eV (see
Figure~\ref{LDA_dos_bands}).
This is perhaps the largest $W$ value among iron arsenides or chalcogenides: BaFe$_2$As$_2$ --- $4.5$
eV~\cite{BaFe2As2_Skornyakov2012,BaFe2As2_Werner2012}, in NaFeAs --- $4.5$ eV~\cite{NaFeAs}, in KFe$_{2}$Se$_2$ --- 3.5
eV~\cite{KFeSeLDADMFT1}.
However, the electron correlation strength is proportional to $U/W$ ratio, which is even larger for BaNi$_2$As$_2$ (0.95) than for BaFe$_2$As$_2$ (0.78).
Thus enlargement of $W$ in BaNi$_2$As$_2$ does not lead to decrease of correlation strength.

It's well known that the Hund's coupling plays a major role in iron-based superconductors~\cite{Haule2009,DeMedici2011}, and it happens because the Fe-3d shell is nearly half-filled.
At half-filling the critical interaction strength $U_c$ (where Mott transition occurs) and, consequently, the quasiparticle renormalization factor $Z$ are larger than those for almost filled band at the same $J$~\cite{DeMedici2011}.
Since here we chose the same Hund's coupling for both Ni-3d and Fe-3d states ($J=0.85$~eV), the correlation strength in BaNi$_2$As$_2$ differs from one in BaFe$_2$As$_2$ mainly due to larger 3d occupancy value.

\begin{figure}[h]
	\includegraphics[width=0.85\linewidth]{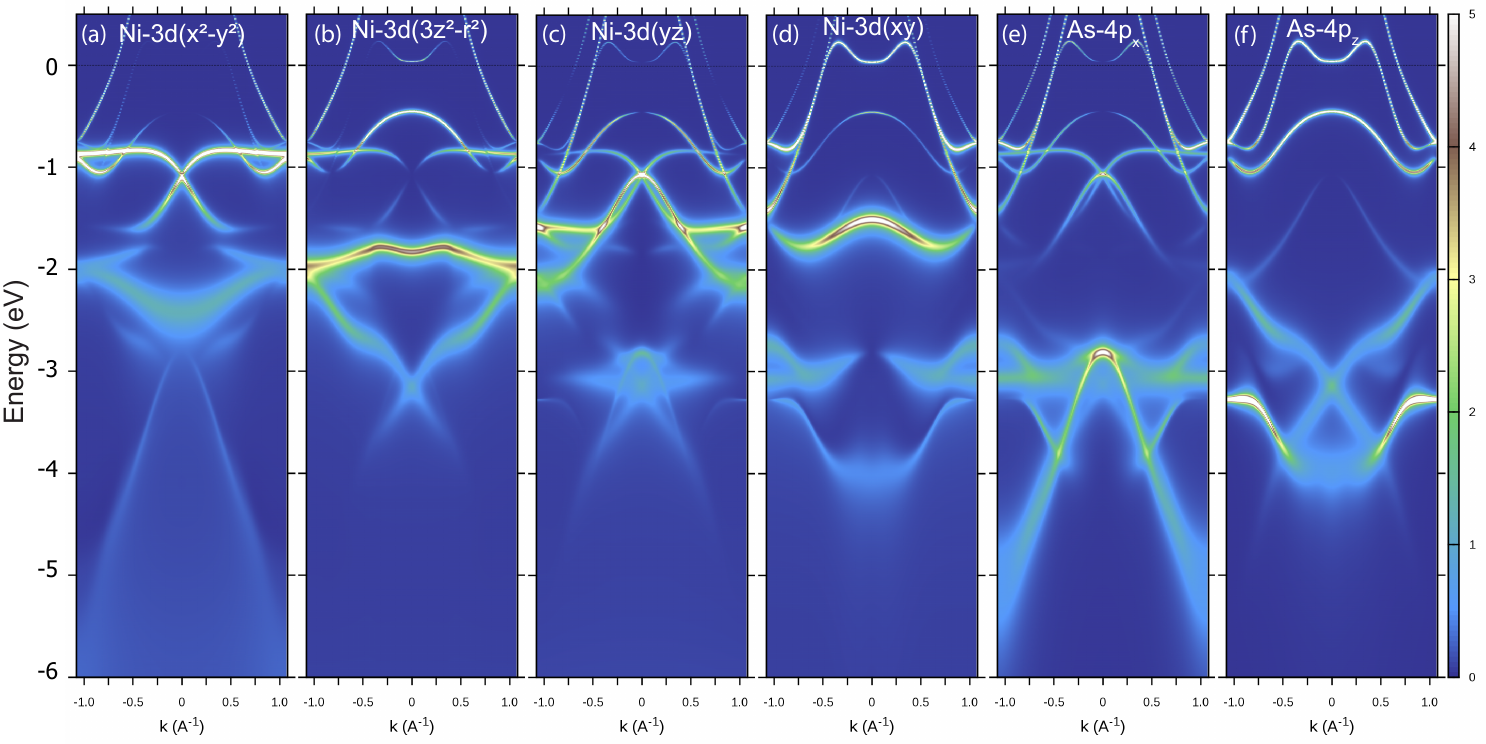}
	\caption{Orbital resolved LDA+DMFT spectral function maps for Ni-3d (a,b,c,d) and As-4p (e,f) states
		in the X-$\Gamma$-X high symmetry direction for BaNi$_2$As$_2$.
		The Fermi level is at zero energy.}
	\label{LDADMFT_orbs}
\end{figure}

Let us emphasize another peculiarity of BaNi$_2$As$_2$ that differs it from regular iron arsenide superconductors.
Once we plot orbital-resolved LDA+DMFT spectral function maps for Ni-3d and As-4p states, one can see that
all these orbitals are present at the Fermi level (see Figure~\ref{LDADMFT_orbs}).
Also orbital resolved spectral function maps are very useful for better understanding of ARPES data.
Main contribution to the spectral function at the Fermi level comes from to Ni-3d$_{xy}$ and As-4p$_z$ states.
The bands around $-3$ eV correspond to Ni-3d$_{x^2-y^2}$, Ni-3d$_{z^2}$ and As-4p states.
The quasiparticle band at $-0.5$ eV near $\Gamma$-point corresponds to the hybrid Ni-3d$_{z^2}$ --- As-4p$_z$ state.
The quasiparticle bands seen below $-2.0$ eV in the ARPES data recoded with horizontal light polarization
(Figure~\ref{LDA_DMFT_ARPES}(c)) represent Ni-3d$_{x^2-y^2}$ and Ni-3d$_{z^2}$ states.
For vertical polarization ARPES data (Figure~\ref{LDA_DMFT_ARPES}(d)) high intensity region around $\Gamma$-point in the range ($-1$; $-2$) eV
belongs to Ni-3d$_{yz,xz}$, Ni-3d$_{xy}$ and Ni-3d$_{z^2}$ states.
The flat band near $-0.8$ eV has Ni-3d$_{x^2-y^2}$ character.
One can see from comparison of Figure~\ref{LDADMFT_orbs} and Figure~\ref{LDA_DMFT_ARPES} that the As-4p states are almost not resolved due to the
suppression of photoemission cross-section and photo-hole lifetime effects --- thus the As-4p states are visible below $-3$ eV as strongly
blurred background.

\section{Conclusions}
In this paper we present comparison of LDA+DMFT results with ARPES data in a wide binding energy range for the BaNi$_2$As$_2$, a close relative of the iron arsenide high temperature superconductors.
The LDA+DMFT calculations for BaNi$_2$As$_2$ show quite weak manifestation of the correlations at the Fermi level, in a very good agreement with ARPES results (as can be seen from direct matching of corresponding band dispersions).

Let us point out that Ni-3d electrons in BaNi$_2$As$_2$ must be treated as essentially correlated ones since $U/W \sim 1$, which means that plane LDA a priori should not work in this case. However LDA+DMFT mass enhancement for BaNi$_2$As$_2$ is found to be only $m^*/m=1.2$ despite the strong Coulomb interactions $U$=4.5~eV and $J$=0.85~eV, in contrast to the iron superconductors with $m^*/m$ ranging from 3 for arsenides up to 5 in chalcogenides~\cite{JTLRev}.
The main reason for weak quasiparticle band renormalization in BaNi$_2$As$_2$ is almost fully occupied Ni-3d states with $9.2$ electrons.
Under the same Hund's coupling increase of the occupation of 3d states provides decrease of the correlation effects in BaNi$_2$As$_2$ w.r.t. BaFe$_2$As$_2$~\cite{DeMedici2011}.
On the other hand, it is shown that, despite the weak manifestation of electron correlations near the Fermi level, the LDA+DMFT calculations capture the remarkable nonlinear reconstruction of the bare LDA bands below $-0.8$~eV due to LDA+DMFT self-energy effects.
Also, the LDA+DMFT self-energy is descriptively explained with the toy model using only the bare DOS as an input.

For realistic comparison between theoretical and experimental results, several details are to be accounted for: the photoemission cross-section, the experimental energy and angular resolutions and the photo-hole lifetime effects. All this provides almost quantitative agreement of the LDA+DMFT spectral function map with ARPES data.

\begin{acknowledgement}
We thank Diamond Light Source synchrotron for access to the beamline I05-ARPES (proposal numbers SI4906 and NT5008) that contributed to the results presented here.
We thank A. Sandvik for making available his maximum entropy program.
We are grateful to Professor E. Z. Kuchinskii and Professor S. V. Streltsov for useful discussion.
This work was supported in part by RFBR grant No. 20-02-00011.
NSP work was also supported in part by the President of Russia grant for young scientists No. MK-1683.2019.2.
The CT-QMC computations were performed at ``URAN'' supercomputer of the Institute of Mathematics and Mechanics of the RAS Ural Branch.
\end{acknowledgement}

\section{Appendix}
\subsection*{Theoretical Spectral Function Maps Including Experimental Details}
\begin{figure*}[!ht]
	\includegraphics[width=0.9\linewidth]{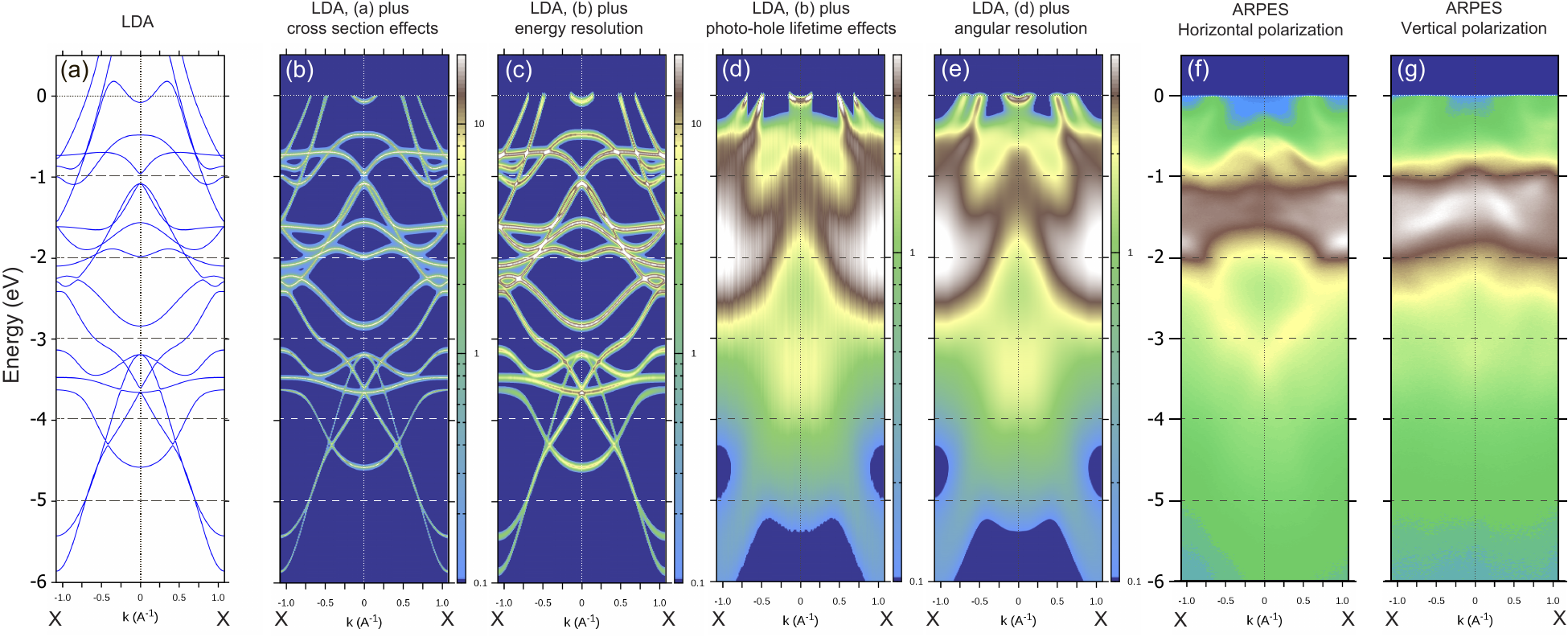}\\
	\includegraphics[width=0.9\linewidth]{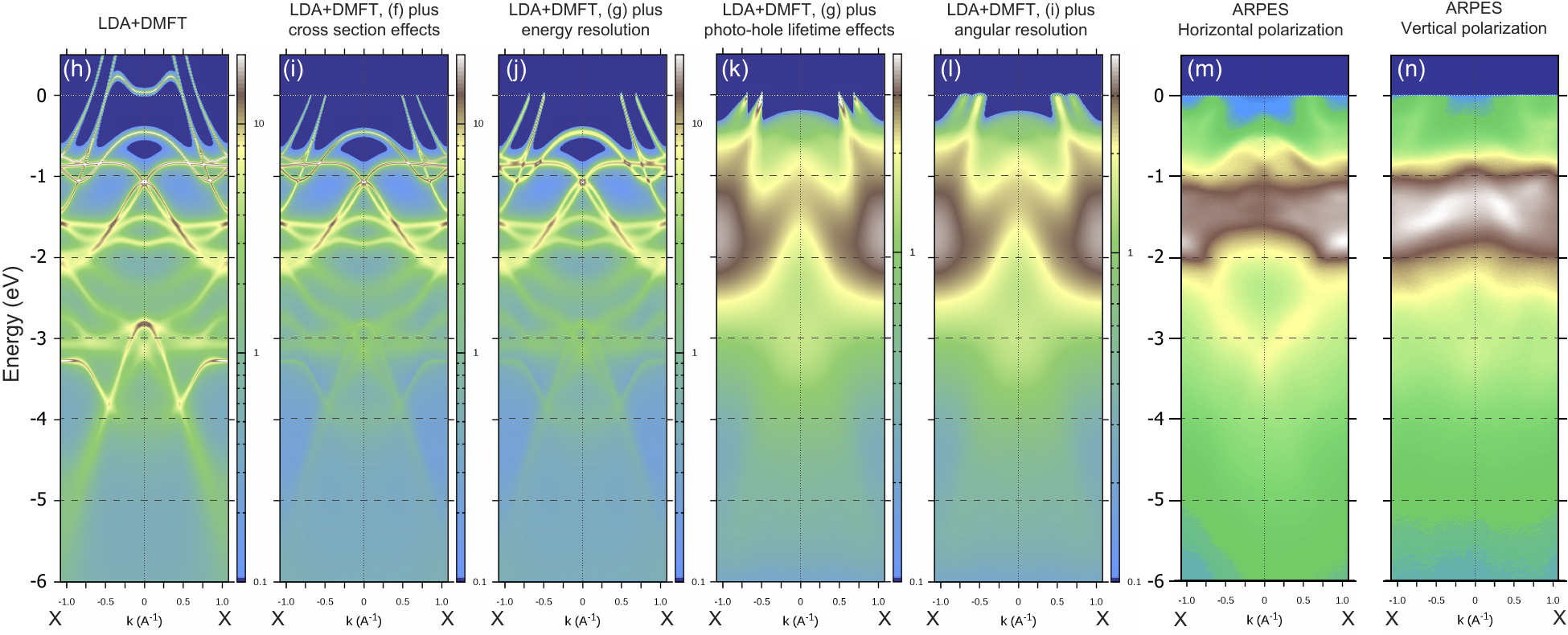}
	\caption{LDA (upper row) and LDA+DMFT (lower row) spectral function maps in the X-$\Gamma$-X high symmetry direction,
		with the photoemission cross-section contributions panels (b,i).
		On the panels (c,j) in addition to (b,i), the energy resolution is taken into account.
		Then on the (d,k) panels the energy resolution and the photo-hole lifetime effects are added.
		Finally on the panels (e,l) the k-space resolution is introduced. Corresponding LDA and LDA+DMFT spectral functions have the same
		intensity scale.
		ARPES data recoding with horizontal (f,m) and vertical (g,n) light polarization.
		The Fermi level is at zero energy.}
	\label{LDA_DMFT_broaden}
\end{figure*}

Here we describe the details about including experimental features~\cite{DMFT_broading}: the photoemission cross-section, the experimental
energy and angular resolutions and the photo-hole lifetime effects.
In Figure~\ref{LDA_DMFT_broaden} on the panels (a) and (h) LDA bands and LDA+DMFT spectral function map are presented in the X-$\Gamma$-X high
symmetry direction.
Next to them on panels (b) and (i) LDA and LDA+DMFT spectral function maps including phoemission cross-section ratio are shown.

For the given experimental photon energy $120$ eV corresponding cross-section ratios can be interploated in the atomic limit as Ni-3d : As-4p :
Ba-5d$\ =5.28:0.06:0.98$ (see~Ref.~\cite{Yeh_Lindau_Tables}).
Also theoretical spectra were multiplied with the Fermi function at experimental temperature 7~K thus
giving quasiparticle bands below the Fermi level.
Because of photoemission cross-section relative amplification, the Ni-3d states become more bright, especially in the ($-3; -2$) eV energy interval.
Then the theoretical spectra obtained above are convoluted with the Gaussian function to take into account the experimental energy resolution
of $20$~meV for LDA (panel c) and LDA+DMFT (panel j).
The effect of the bare LDA band broadening due to experimental resolution is quite noticeable (Figure~\ref{LDA_DMFT_broaden}(c)).
But for the LDA+DMFT bands brodeaning due to experimental resolution is hardly observable because electron correlations themselves produce
significant band damping.

In our earlier work~Ref.~\cite{Sr2RuO4_broadening_Gaussian_2007} it was demonstrated that an energy dependent broadening of the theoretical
spectral functions gives better agreement with the experimental data.
To this end theoretical quasiparticle bands were convoluted with a Gaussian with a full width at
half maximum increasing as $C\cdot\epsilon_B+\Gamma_{exp}$. Here $\epsilon_B$ is the binding energy, $\Gamma_{exp}$ is the experimental
resolution, and $C$ characterizes the increase of the broadening with energy upon moving
away from Fermi level due to photo-hole lifetime effects (for more details
see~Refs.~\cite{Sr2RuO4_broadening_Gaussian_2007,broadening_Lorentzian_1992,broadening_Gaussian_2000}).
The maximum allowable broadening was limited to 0.5~eV for the energy lower than $-1$~eV.
Corresponding broadened LDA and LDA+DMFT spectral function maps
are drawn on the panels (d) and (k) of the Figure~\ref{LDA_DMFT_broaden}.
The photo-hole lifetime effects provide
rather strong damping of the quasiparticle bands making them nearly indistinguishable.
Maximum value of the spectral function between panels c and d (j and k) drops down about 5 times.
Thus to keep quasiparticle bands visible enough, we change corresponding intensity scales (see color scale tick labels).
Finally to complete realistic comparison of the theory with ARPES we consider the experimental angular resolution of $0.5$ degrees
(k-space resolution $0.04\ \AA^{-1}$) by the Gaussian function for LDA (panel e) and LDA+DMFT (panel l).

\begin{figure}[h]
	\includegraphics[width=0.85\linewidth]{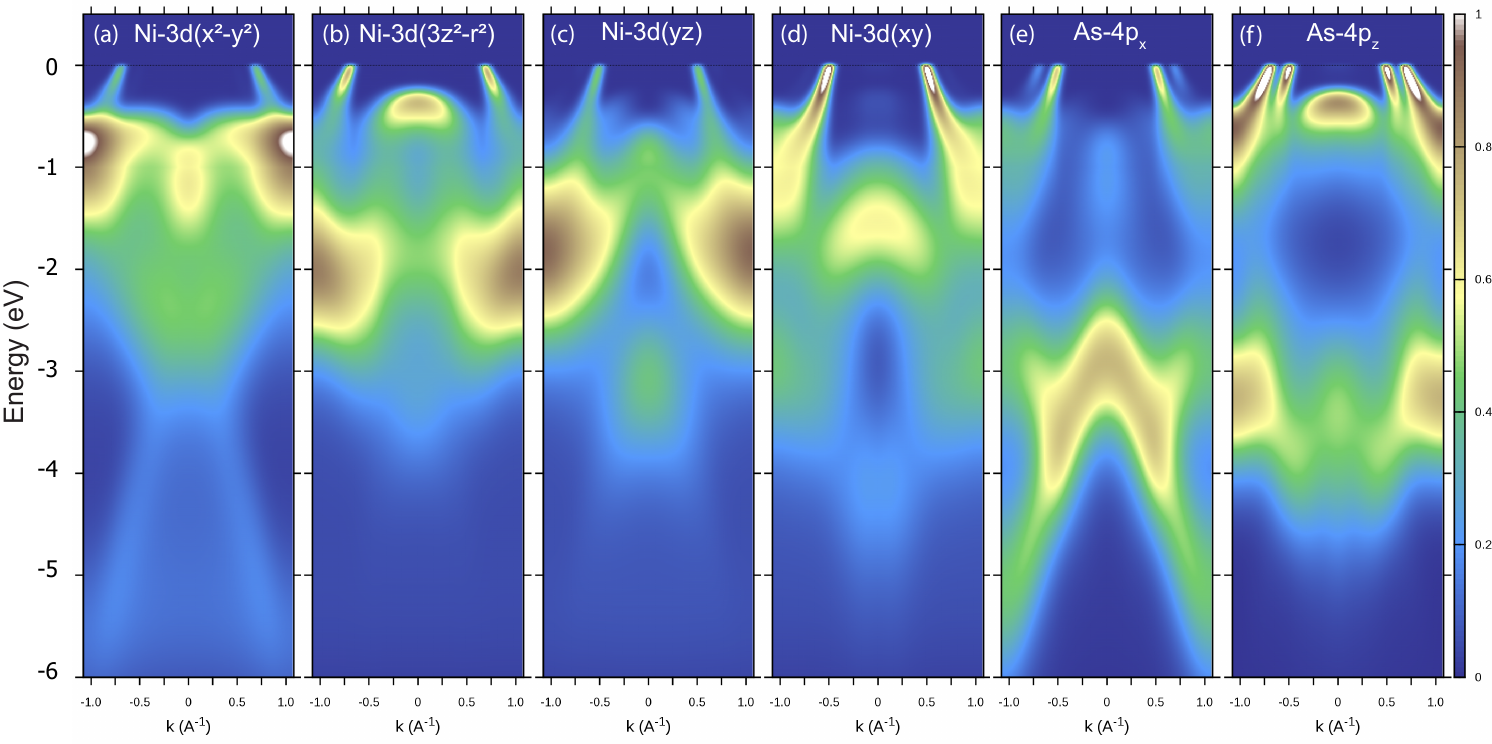}
	\caption{Orbital resolved LDA+DMFT spectral function maps with taking into account the experimental resolution and
	the photo-hole lifetime effects for Ni-3d (a,b,c,d) and As-4p (e,f) states
	in the X-$\Gamma$-X high symmetry direction for BaNi$_2$As$_2$.
	The Fermi level is at zero energy.}
	\label{LDADMFT_orbs_broaden}
\end{figure}

\subsection*{Theoretical PES}
Theoretical PES (Figure~\ref{PES}) was obtained by integrating the LDA+DMFT spectral function along the cut passing through the Brillouin zone
diagonal (X-$\Gamma$-X direction) for BaNi$_2$As$_2$ and over the entire Brillouin zone for BaFe$_2$As$_2$.
The experimental features included for the theoretical PES were the same as above, but without angular resolution.

\subsection*{Orbital Resolved Theoretical Spectral Function Maps}
To show contribution of bands of different symmetry the orbital resolved LDA+DMFT spectral function maps (Ni-3d and As-4p states) with
experimental details included are presented in Figure~\ref{LDADMFT_orbs_broaden}.

\subsection*{Photon Energy Dependence of ARPES}
\begin{figure}[h]
	\includegraphics[width=0.6\linewidth]{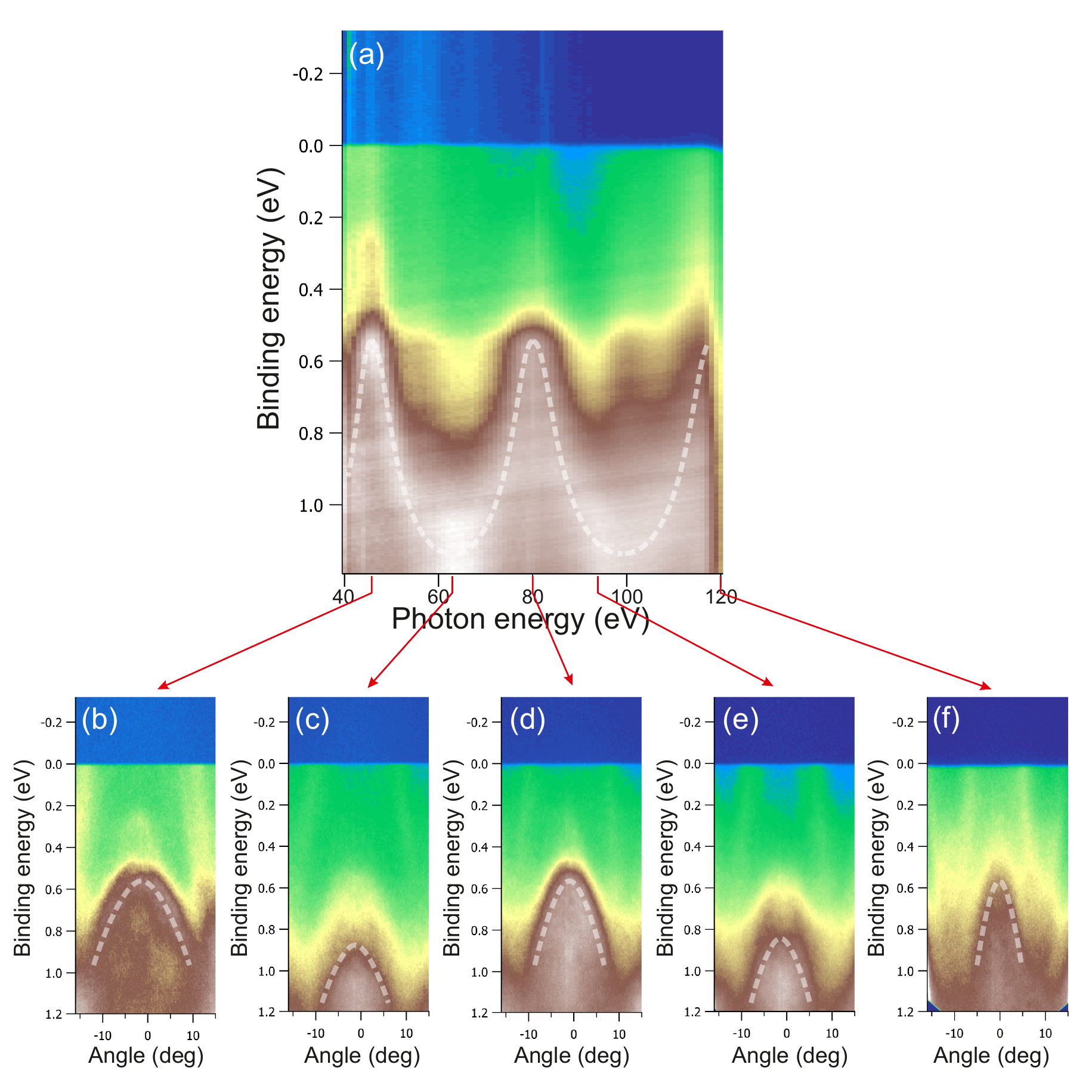}
	\caption{(a) Photon energy dependence of the spectrum corresponding to the normal emission, $k_x=0$ $k_y=0$. Individual energy-momentum cuts are shown below:
		(b) 45~eV, (c) 62~eV, (d) 80~eV, (e) 95~eV, (f) 120~eV.
	Apparent narrowing of the band in the horizontal dimension when going from (b) to (f) is connected to the squeezing of the momentum space in angular representation with the increase of the kinetic energy of the emitted electron.}
	\label{ARPES_kz}
\end{figure}

The h$\nu$-dependent data set imaged with the photon energy ranging from 40 to 120~eV is presented in Figure~\ref{ARPES_kz}. The dispersion of the band ascending up to the binding energy of about 0.5~eV at the very center of the Brillouin zone ($k_x=0$, $k_y=0$), can be seen.  The observed quasiperiodic variation of the intensity distribution as a function of the photon energy is inline with the expected cycling of the $k_z$ value dominating the spectrum. From considering the calculated band dispersion (Fig.~\ref{LDA_dos_bands}(b), Fig.~\ref{ARPES_LDADMFT_bands}(a)), we know that the lowest binding energy for the band in question is achieved at $k_z=\pi$. Consequently, the measured variation of the spectrum with photon energy provides independent evidence for the fact that 120~eV corresponds in photoemission to the $k_z$ values close to $\pi$.
We note that the apparent level of the spectra quality and resolution in this photon energy scan, and in the energy-momentum cuts along the momentum parallel to the sample surface, are substantially different. The selection of the photon energy of 120~eV, used for recording the high-statistics data (Fig.~\ref{ARPES_LDADMFT_bands}(b,c)), was determined by the general clarity of the spectrum and superior discernibility of the band dispersion. 

\subsection*{Toy Model for the Self-Energy Calculation}
In this paper to account for correlation interactions we did LDA+DMFT calculation for BaNi$_2$As$_2$.
Strictly speaking DMFT is a way of complete summation of full set of diagrams (or in other words -- electron scattering processes) for the case of Hubbard (local) Coulomb interaction. For such kind of exact theories
it is difficult to extract, so to say, leading contributions which are important for simple discriptive  visualisations and interpretation of obtained results. 

Thus, in this section we propose a toy model allowing us to pick up the ``leading'' (as we suppose) electron-electron scattering processes to the noninteracting system.
This toy model, as we show below, captures difference of the correlation strength between BaFe$_2$As$_2$ and BaNi$_2$As$_2$ and provide qualitative explanation of the DMFT self-energy in a wide window around Fermi level.

The \emph{physical} idea we will rely upon is rather straightforward: for each electronic state we will calculate the scattering rate resulting from the particle-particle interaction. In terms of the Green function formalism such scattering rate corresponds to the imaginary part of the self-energy, $\Sigma^{''}$.

\begin{figure}[h]
	\includegraphics[width=0.5\linewidth]{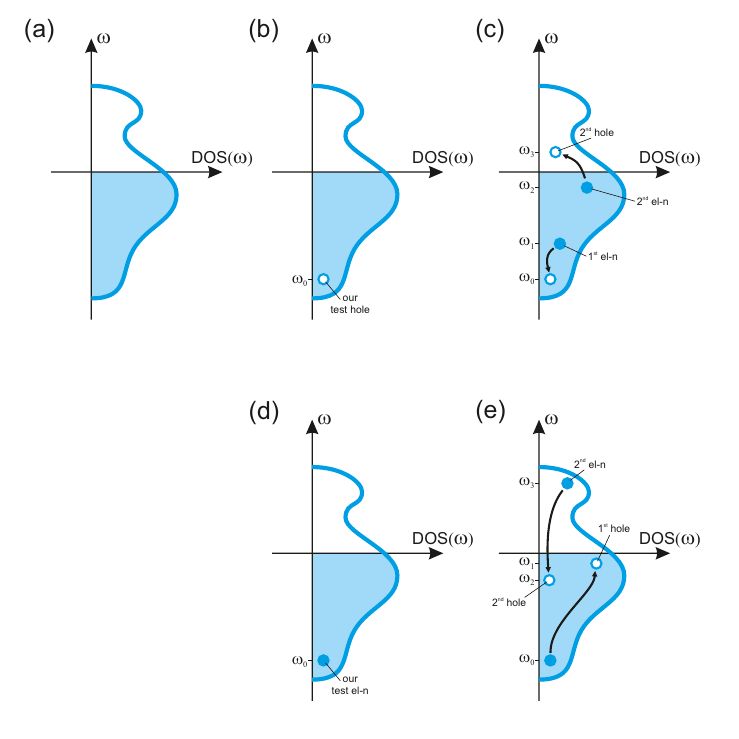}
	\caption{Illustration of the general two-particle processes involving either an electron or a hole in a given state, characterized by its energy $\omega_0$.  (a) Exemplar density of states, filled states are marked by the light blue. (b) A test hole residing at
		$\omega_0$. (c) An electron-electron scattering process resulting in the filling of the test hole. (d) A test electron at $\omega_0$. (e) A
		process resulting in the scattering of the test electron. Note that the actual scattering event depicted here has vanishingly low probability.
		However, such electron scattering will start to provide a contribution to the overall lifetime at elevated temperatures, and will become
		dominating once $\omega_0$ will change sign.}
	\label{smplmodel}
\end{figure}
Let's consider a hole occurred in the filled part of the conduction band [Figure~\ref{smplmodel}(a,b)]. The two-particle processes resulting in
the filling of this hole are depicted in the panel (c). As soon as we consider the
effects resulting from the particle-particle interaction, the energy in this process should be conserved, $\omega_1+\omega_2 = \omega_0 + \omega_3$. Upon such unitary electron-electron scattering, the momentum is also conserved. However, we will omit the momentum in our consideration
altogether, and will characterise the states only by their energy.

The probability of filling our test hole in a unitary process shown in Figure~\ref{smplmodel}(c) is than proportional to the probability to (i)
find an electron at $\omega_1$, to (ii) find an electron at $\omega_2$, and to (iii) find a hole at $\omega_3$:
\begin{equation}
  {\rm DOS}(\omega_1)f(\omega_1)\cdot {\rm DOS}(\omega_2)f(\omega_2) \cdot {\rm DOS}(\omega_3)[1-f(\omega_3)],
\end{equation}
where ${\rm DOS}(\omega)$ is the density of states, and $f(\omega)$ is the Fermi-Dirac distribution. Generally there should also be a matrix
element of the corresponding transition\,---\,a product of the initial and final electronic wave functions over the particle-particle
potential\,---\,but we will assume it to be equal for all possible transitions.

The probability for the test hole to scatter is expressed then as an integral over all possible unitary processes.
Taking into account that $\omega_3 = \omega_1 + \omega_2 - \omega_0$ owing to the energy conservation,
\begin{equation}
\Sigma^{''}_{\rm hole}(\omega_0) \sim  \int\limits_{\omega_1}\int\limits_{\omega_2}{\rm DOS}(\omega_1)f(\omega_1)\cdot
\end{equation}
\begin{equation}
\cdot {\rm DOS}(\omega_2)f(\omega_2) \cdot {\rm DOS}(\omega_1 + \omega_2 - \omega_0)[1-f(\omega_1 + \omega_2 - \omega_0)]{\rm d}\omega_1{\rm
d}\omega_2.
\end{equation}
In order to complete the consideration of all possible processes involving the state $\omega_0$, we formally need to consider the possible
scattering of the test electron placed into this state, see Figure~\ref{smplmodel}(d,e). Similarly to the case of the test hole, this probability
is expressed as
 \begin{equation}
\Sigma^{''}_{\rm el-n}(\omega_0) \sim  \int\limits_{\omega_1}\int\limits_{\omega_2}{\rm DOS}(\omega_1)[1-f(\omega_1)]\cdot
\end{equation}
\begin{equation}
\cdot {\rm DOS}(\omega_2)[1-f(\omega_2)] \cdot {\rm DOS}(\omega_1 + \omega_2 - \omega_0)f(\omega_1 + \omega_2 - \omega_0){\rm d}\omega_1{\rm
d}\omega_2,
\end{equation}
and the imaginary part of the self-energy is $\Sigma^{''}(\omega_0) = \Sigma^{''}_{\rm hole}(\omega_0) + \Sigma^{''}_{\rm el-n}(\omega_0).$ As the self-energy is an analytic function, the real part of the self energy can be restored from $\Sigma^{''}$ via the Kramers-Kronig
transform.

\begin{figure}[h]
	\includegraphics[width=0.7\linewidth]{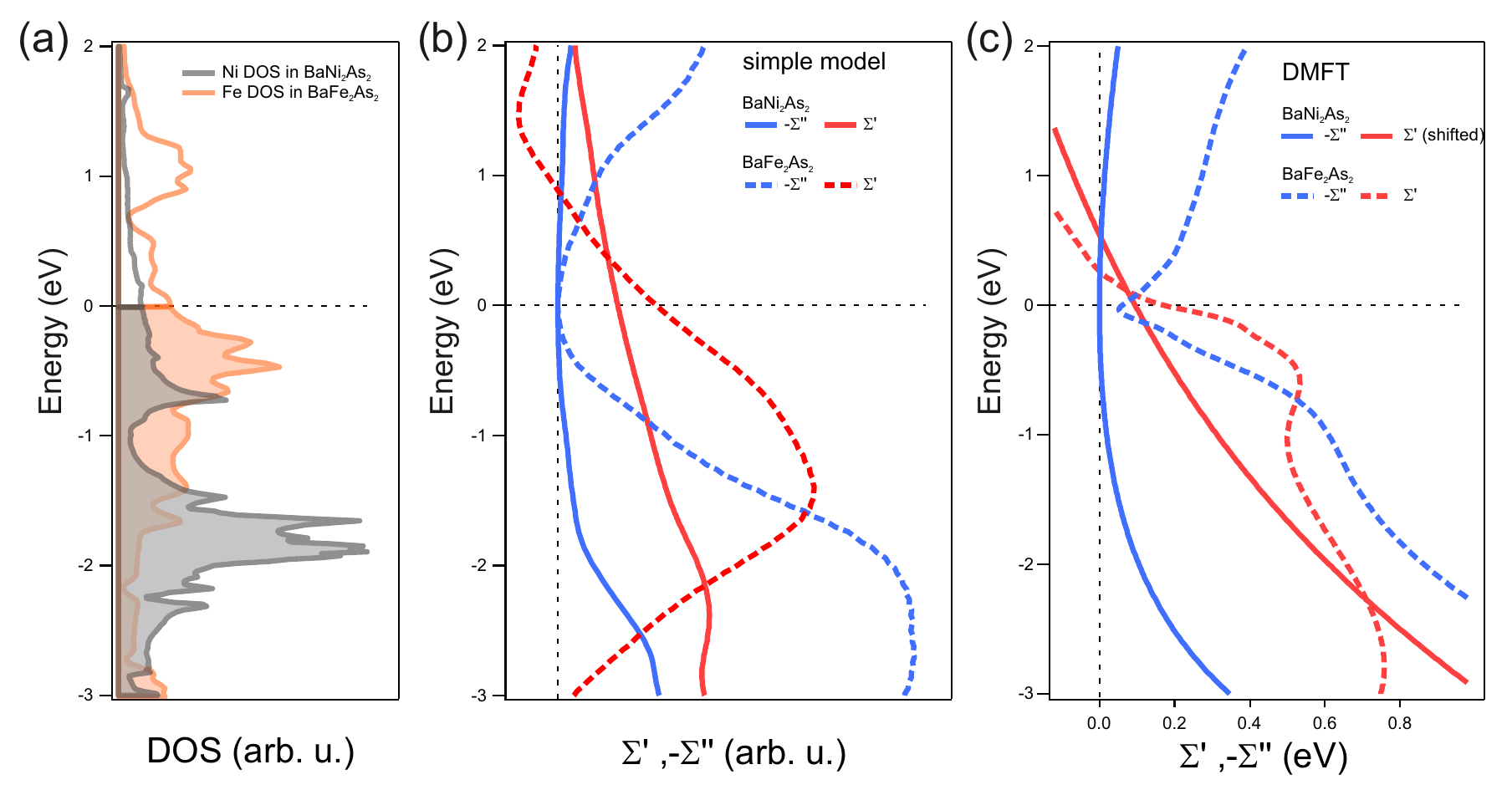}
	\caption{(a) Density of states related to the transition metal 3d orbitals in BaNi$_2$As$_2$ and in BaFe$_2$As$_2$. (b) Self-energy derived
within the simple model described in this section. Matrix element of all unitary transitions is assumed to be equal, and also \emph{equal
between the two compounds}. (c) The self-energy from the DMFT calculation, average for all orbitals. Note that $\Sigma^{'}$ for the Ni-based compound was shifted along the vertical axis.}
	\label{Selfencompar}
\end{figure}

In the Figure~\ref{Selfencompar} we present the results of the self-energy calculations according to this simple scheme, using the LDA-derived density of states related to the $3d$ ions in BaNi$_2$As$_2$ and in BaFe$_2$As$_2$ as an input, and compare it to the self-energy calculated by the DMFT.
The DMFT self-energy doesn't exhibit large orbital variation, the presented curves correspond to the values averaged for the all 3d orbitals.
For simplicity we assumed the temperature to be much smaller than the other energy scales involved\,---\,it seems to be a reasonable approximation here, as characteristic band energies are of the order of electron-Volts.
Although there are noticeable differences for the two ways of the self-energy calculation, the drastic change of the self-energy behavior in the vicinity of the Fermi level between the Fe- and Ni-based compounds, is well captured even in the simple model.
One of the most important parameters that one can look at, is the derivative of $\Sigma^{'}$ at the Fermi level, which for the Ni compound appears to be 4 times smaller than for Fe one in the DMFT calculation, and 5 times smaller in the presented simple model.

The calculation scheme described in this section can be regarded as a mnemonic rule with a certain physical meaning.
The toy model results support the idea that the band renormalization at the Fermi level is largely determined by the bare density of $3d$ states and
the $3d$ band filling. In particular, the DOS$(\omega=0)$ and large amount of both occupied and empty states are important parameters. Finally, these simple considerations illustrate weakness of correlation effect in  BaNi$_2$As$_2$ in contrast to BaFe$_2$As$_2$ in agreement with LDA+DMFT and ARPES data.

\bibliography{bib_file}

\end{document}